\documentclass[twocolumn,showpacs,preprintnumbers,amsmath,amssymb]{revtex4}

\usepackage{graphicx}
\usepackage{dcolumn}
\usepackage{bm}

\begin{document}

\preprint{}

\title{Experimental observation of the spin Hall effect of light on
a nano-metal film via weak measurements}
\author{Xinxing Zhou}
\author{Zhicheng Xiao}
\author{Hailu Luo}\email{hailuluo@hnu.edu.cn}
\author{Shuangchun Wen}\email{scwen@hnu.edu.cn}
\affiliation{Key Laboratory for Micro-/Nano-Optoelectronic Devices
of Ministry of Education, College of Information Science and
Engineering, Hunan University, Changsha 410082, People's Republic of
China}
\date{\today}

\begin{abstract}
We theorize the spin Hall effect of light (SHEL) on a nano-metal
film and demonstrate it experimentally via weak measurements. A
general propagation model to describe the relationship between the
spin-orbit coupling and the thickness of the metal film is
established. It is revealed that the spin-orbit coupling in the SHEL
can be effectively modulated by adjusting the thickness of the metal
film, and the transverse displacement is sensitive to the thickness
of metal film in certain range for horizontal polarization light.
Importantly, a large negative transverse shift can be observed as a
consequence of the combined contribution of the ratio and the phase
difference of Fresnel coefficients.
\end{abstract}


\maketitle

\section{Introduction}\label{SecI}
The spin Hall effect of light (SHEL) manifests itself as the split
of a linearly polarized beam into left- and right-circular
components when a beam propagates through homogeneous media. The
splitting in the SHEL, governed by the angular momentum conservation
law~\cite{Onoda2004,Bliokh2006}, takes place as a result of an
effective spin-orbit coupling. The SHEL is sometimes referred to as
the Imbert-Fedorov effect, which was theoretically predicted by
Fedorov and experimentally confirmed by
Imbert~\cite{Fedorov1965,Imbert1972}. Generally the transverse shift
of the SHEL is on the subwavelength scale, and it is difficult to be
directly measured with the conventional experimental methods. In
2008, benefiting from the weak measurement technique, Hosten and
Kwiat first measured the transverse displacement of refracted
light~\cite{Hosten2008}. The SHEL has been widely investigated in
different physical systems, such as high-energy
physics~\cite{Gosselin2007,Dartora2011},
plasmonics~\cite{Gorodetski2008,2Bliokh2008,Vuong2010}, optical
physics~\cite{Bliokh2008,Aiello2008,Haefner2009,Herrera2010,Qin2009,Luo2009},
and semiconductor physics~\cite{Menard2010,Menard2009}.

The SHEL holds great potential applications, such as manipulating
electron spin states and precision metrology~\cite{Hosten2008}. The
SHEL itself may become a useful metrological tool for characterizing
the refractive index variations of nanostructure. Thus, the
relationship between SHEL and nanostructure is important, yet it is
not fully understood. To measure the refractive index variations at
subwavelength scale, we need to establish the relationship between
the SHEL and the structural parameters of the nanostructure. It is
well known that the SHEL manifests itself as the spin-orbit
coupling. Now a question arises: What role does the structural
parameters of the subwavelength nanostructure play in the spin-orbit
coupling?

In this work, we establish a general propagation model to describe
the SHEL on the nano-metal film and reveal the impact of the
structural parameters on the SHEL. We find that the spin-orbit
coupling in the SHEL can be effectively modulated by adjusting the
thickness of the metal film. It should be noted that the interesting
SHEL on this structure is different from that on pure glass
prism~\cite{Hosten2008,Qin2009}, metal bulk~\cite{Hermosa2011}, and
layered nanostructure~\cite{2Luo2011}. The paper is organized as
follows. First, we analyze the SHEL on the nano-metal film
theoretically. Our findings indicate that the transverse
displacement of the SHEL is sensitive to the thickness of the metal
film and undergoes a large negative value for horizontal
polarization. Next, we focus our attention on the experiment (weak
measurements). Here, the sample is a BK7 substrate coated with a
thin Ag film. The experimental results are in good agreement with
the theory. Finally, a conclusion is given in the fourth section.

\begin{figure}
\includegraphics[width=8.5cm]{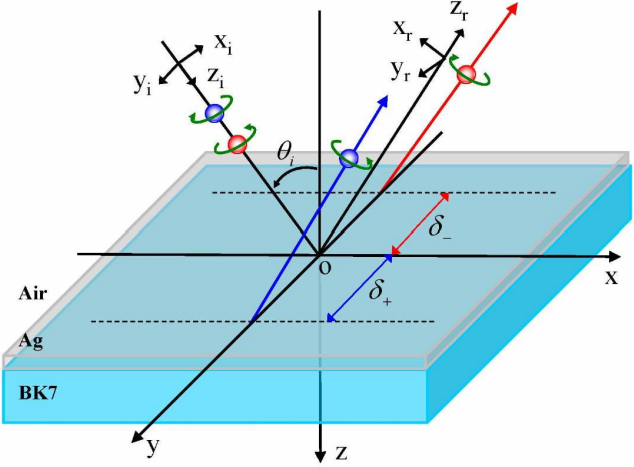}%
\caption{\label{Fig1} (Color online) Schematic of SHEL on a
nano-metal film. A linearly polarized beam reflects on the model
composed of air, Ag film and the BK7 glass substrate and then splits
into left- and right-circularly polarized light, respectively.
$\delta_{+}$ and $\delta_{-}$ indicate the transverse shift of left-
and right-circularly polarized components. Here, $\theta_{i}$ is the
incident angle and the Goos-H\"{a}nchen shift is not considered.}
\end{figure}

\section{Theoretical model}\label{SecII}
Figure~\ref{Fig1} schematically illustrates the SHEL of beam
reflection on a nano-metal film in Cartesian coordinate system. The
$z$ axis of the laboratory Cartesian frame ($x,y,z$) is normal to
the interface of the metal film at $z=0$. The incident and reflected
electric fields are presented in coordinate frames ($x_i,y_i,z_i$)
and ($x_r,y_r,z_r$), respectively. In the spin basis set, the
angular spectrum can be written as
$\tilde{\mathbf{E}}_i^H=(\tilde{\mathbf{E}}_{i+}+\tilde{\mathbf{E}}_{i-})/{\sqrt{2}}$
and
$\tilde{\mathbf{E}}_i^V=i(\tilde{\mathbf{E}}_{i-}-\tilde{\mathbf{E}}_{i+})/{\sqrt{2}}$.
Here, $H$ and $V$ represent horizontal and vertical polarizations,
respectively. The positive and negative signs denote the left- and
right-circularly polarized (spin) components, respectively.

The incident monochromatic Gaussian beam can be formulated as a
localized wave packet whose spectrum is arbitrarily narrow, and can
be written as
\begin{equation}
\widetilde{\mathbf{E}}_{i\pm}=(\mathbf{e}_{ix}+i\sigma\mathbf{e}_{iy})\frac{w_{0}}{\sqrt{2\pi}}\exp
\left[-\frac{w_{0}^{2}(k_{ix}^{2}+k_{iy}^{2})}{4}\right]\label{3},
\end{equation}
where $w_{0}$ is the beam waist. The polarization operator
$\sigma=\pm1$ corresponds to left- and right-circularly polarized
light, respectively. According to the transversality, we can obtain
the reflected field~\cite{2Luo2011}
\begin{eqnarray}
\left[\begin{array}{cc}\widetilde{\mathbf{E}}_{r}^{H} \\
\widetilde{\mathbf{E}}_{r}^{V}\end{array}\right]=\left[
\begin{array}{cc}
r_p &\frac{k_{ry} \cot\theta_i(r_p+r_s) }{k_0} \\
-\frac{k_{ry} \cot\theta_i(r_p+r_s) }{k_0} & r_s
\end{array}\right] \left[\begin{array}{cc}\widetilde{\mathbf{E}}_{i}^{H} \\
\widetilde{\mathbf{E}}_{i}^{V}\end{array}\right]\label{4}.
\end{eqnarray}
Here, $r_{p}$ and $r_{s}$ denote Fresnel reflection coefficients for
parallel and perpendicular polarizations, respectively. $k_{0}$ is
the wave number in free space.

From Eqs.~(\ref{3}) and~(\ref{4}), we can obtain the expressions of
the reflected angular spectrum
\begin{equation}
\widetilde{\mathbf{E}}_{r}^{H}=\frac{r_{p}}{\sqrt{2}}\left[\exp(+ik_{ry}\delta_{r}^{H})\widetilde{\mathbf{E}}_{r+}+\exp(-ik_{ry}\delta_{r}^{H})\widetilde{\mathbf{E}}_{r-}\right]\label{5},
\end{equation}
\begin{equation}
\widetilde{\mathbf{E}}_{r}^{V}=\frac{ir_{s}}{\sqrt{2}}\left[-\exp(+ik_{ry}\delta_{r}^{V})\widetilde{\mathbf{E}}_{r+}+\exp(-ik_{ry}\delta_{r}^{V})\widetilde{\mathbf{E}}_{r-}\right]\label{6}.
\end{equation}
Here, $\delta_{r}^{H}=(1+r_{s}/r_{p})\cot\theta_{i}/k_{0}$,
$\delta_{r}^{V}=(1+r_{p}/r_{s})\cot\theta_{i}/k_{0}$, and
$\widetilde{\mathbf{E}}_{r\pm}$ can be written as
\begin{equation}
\widetilde{\mathbf{E}}_{r\pm}=(\mathbf{e}_{rx}+i\sigma\mathbf{e}_{ry})\frac{w_{0}}{\sqrt{2\pi}}\exp
\left[-\frac{w_{0}^{2}(k_{rx}^{2}+k_{ry}^{2})}{4}\right]\label{7}.
\end{equation}
It is known that the spin-orbit coupling is the intrinsic physical
mechanism of the SHEL. We note that, in Eqs.~(\ref{5})
and~(\ref{6}), the terms $\exp(\pm ik_{ry}\delta_{r}^{H})$ and the
$\exp(\pm ik_{ry}\delta_{r}^{V})$ indicate the spin-orbit coupling
terms in the case of horizontal and vertical
polarizations~\cite{Hosten2008}. The spin-orbit coupling terms stem
from the transverse nature of the photon polarization: The
polarizations associated with the plane-wave components undergo
different rotations in order to satisfy the transversality after
reflection~\cite{Hosten2008}. We can find that increasing or
decreasing term $\delta_{r}^{H,V}$ will significantly enhance or
suppress the spin-orbit coupling effect.

It is noted that the real parts of the spin-orbit coupling terms
$\delta_{r}^{H,V}$ denote the spatial shift of the
SHEL~\cite{Aiello2009}. Hence, we can obtain the initial transverse
displacement of the SHEL on the nano-metal film:
\begin{equation}
\delta_{\pm}^{H}=\mp\frac{\lambda}{2\pi}\left[1+\frac{|r_{s}|}{|r_{p}|}\cos(\varphi_{s}-\varphi_{p})\right]\cot\theta_{i}\label{8},
\end{equation}
\begin{equation}
\delta_{\pm}^{V}=\mp\frac{\lambda}{2\pi}\left[1+\frac{|r_{p}|}{|r_{s}|}\cos(\varphi_{p}-\varphi_{s})\right]\cot\theta_{i}\label{9},
\end{equation}
where $r_{p,s}=|r_{p,s}|\exp(i\varphi_{p,s})$ and $\lambda$ is
wavelength of the incident beam. Calculating the reflected shifts of
the SHEL requires the explicit solution of the boundary conditions
at the interfaces. Thus, we need to know the generalized Fresnel
reflection of the metal film,
\begin{eqnarray}
r_{A}=\frac{R_{A}+R_{A}^{'}\exp(2ik_{0}\sqrt{\varepsilon-\sin^{2}\theta_{i}}d)}{1+R_{A}R_{A}^{'}\exp(2ik_{0}\sqrt{\varepsilon-\sin^{2}\theta_{i}}d)}.
\end{eqnarray}
Here, $A\in\{p,s\}$, $R_{A}$ and $R_{A}^{'}$is the Fresnel
reflection coefficients at the first interface and second interface,
respectively. $\varepsilon$ and $d$ represent the permittivity and
thickness of the metal film, respectively.

\begin{figure}
\includegraphics[width=8.5cm]{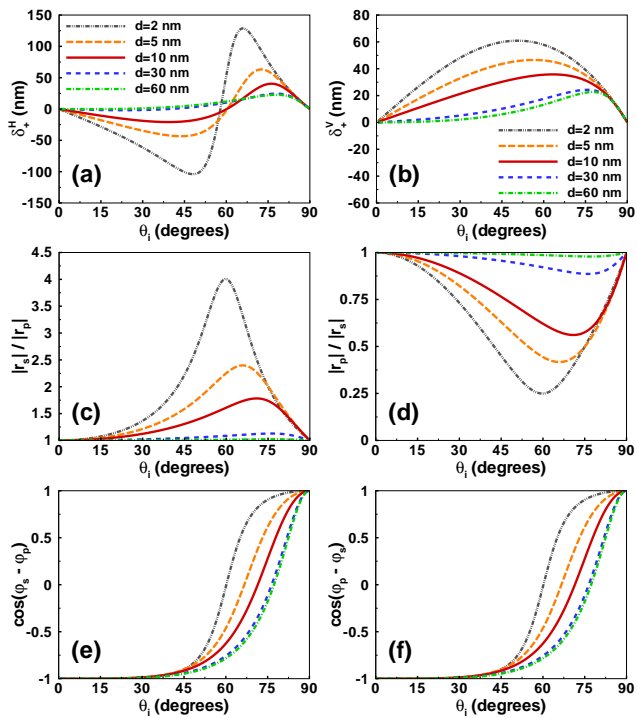}
\caption{\label{Fig2} (Color online) Role of the thickness of
nano-meta film in the SHEL. (a) and (b) represent the transverse
displacements of the SHEL on the thin Ag film under the condition of
horizontal and vertical polarization. We choose the thickness of the
thin metal film from 2 to $60\mathrm{nm}$. (c) and (d) show the
value of $|r_{s}|$/$|r_{p}|$ and $|r_{p}|$/$|r_{s}|$. (e) and (f)
denote the value of $\cos(\varphi_{s}-\varphi_{p})$ and
$\cos(\varphi_{p}-\varphi_{s})$ for the different thickness. Here,
the permittivity of Ag is chosen as $\varepsilon=-18+0.5i$ and the
refractive index of the BK7 substrate is chosen as $n=1.515$ at
$632.8\mathrm{nm}$. }
\end{figure}

To obtain a clear physical picture, we plot Fig.~\ref{Fig2} to
reveal what role the thickness of the nano-meta film plays in the
spin-orbital coupling. Figure~\ref{Fig2}(a) and~\ref{Fig2}(b) show
the initial transverse shifts of the SHEL with different film
thickness. In the case of horizontal polarization, we find that the
transverse displacement is extremely sensitive to the thickness when
it is less than about $10\mathrm{nm}$. We find that this interesting
phenomenon is attributed to the large variations of
$|r_{s}|/|r_{p}|$ [Fig.~\ref{Fig2}(c)]. However, as for vertical
polarizations, the transverse shift is insensitive to the thickness
because of small variations of $|r_{p}|/|r_{s}|$
[Fig.~\ref{Fig2}(d)]. It should be noted that, from Eqs.~(\ref{5})
and~(\ref{8}), the term of $|r_{s}|/|r_{p}|$ plays a dominant role
in spin-orbit coupling. Hence, we can enhance or suppress the SHEL
effectively by modulating the thickness of the metal film. Similar
effect can also be observed in a layered nanostructures, in which
the transverse displacement changes periodically with the air gap
increasing or decreasing due to the optical Fabry-Perot
resonance~\cite{2Luo2011}.

In the case of horizontal polarization, the transverse shift
experiences large negative value [Fig.~\ref{Fig2}(a)], which is
different from the SHEL on a metal bulk~\cite{Hermosa2011}. From
Eqs.~(\ref{8}) and~(\ref{9}), we can find that, for a fixed incident
angle, negative shifts entail the combined contributions of the
large ratio of Fresnel coefficients ($|r_{s}|/|r_{p}|$ or
$|r_{p}|/|r_{s}|$) and phase difference induced negative
$\cos(\varphi_{s}-\varphi_{p})$ or $\cos(\varphi_{p}-\varphi_{s})$
[Fig.~\ref{Fig2}(e) and~\ref{Fig2}(f)] which are due to the material
properties of the metal film. We conclude that large negative
transverse displacement only exists in the case of horizontal
polarization while always is positive under the condition of
vertical polarizations. It is indicated that by rotating the
polarization of incident light beam, we are able to switch the
direction of the spin accumulation~\cite{Luo2011} effectively.
Similar phenomena also occur in electronic system. Here, the spin
accumulation can be switched by altering the directions of an
external magnetic field~\cite{Sinova2004,Kimura2007,Mihaly2008}. By
rotating the polarization plane of the exciting light, the
directions of spin current can be switched in a semiconductor
microcavity due to the spin Hall
effect~\cite{Kavokin2005,Leyder2007}.

\section{Experimental observation}\label{SecIII}
To detect the tiny transverse shifts, we use the signal enhancement
technique known as the weak
measurements~\cite{Aharonov1988,Ritchie1991}. Note that the weak
measurements has attracted a lot of attention and holds great
promise for precision
metrology~\cite{Pryde2005,Dixon2009,Brunner2010,Kocsis2011,Feizpour2011,Zilberberg2011}.
The theoretical analysis of the SHEL on nano-metal film has yielded
two major results: sensitive SHEL in extremely thin metal film and
large negative beam shift of horizontal polarized incident beam.
However, we inevitably face a major obstacle that prevents us from
experimentally corroborating the first claim because fabricating Ag
film thinner than $10\mathrm{nm}$ would unavoidably involve large
technical errors. Nonetheless, we still attempt to verify the
validity of our theory by measuring the SHEL in large film
thickness. In this section, we choose the BK7 glass substrate coated
Ag film as our sample (with three different thickness
$10\mathrm{nm}$, $30\mathrm{nm}$ and $60\mathrm{nm}$.

\begin{figure}
\includegraphics[width=9cm]{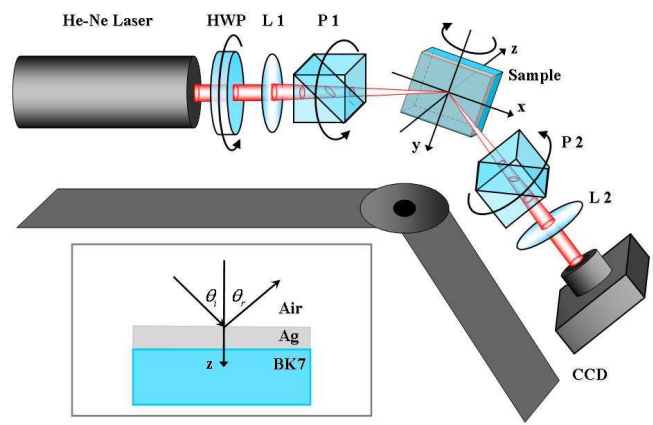}
\caption{\label{Fig3} (Color online) Experimental setup: Sample, a
BK7 glass substrate coated with Ag film. L1 and L2, lenses with
effective focal length $50\mathrm{mm}$ and $250\mathrm{mm}$,
respectively. HWP, half-wave plate (for adjusting the intensity). P1
and P2, Glan Laser polarizers. CCD, charge-coupled device (Coherent
LaserCam HR). The light source is a $17\mathrm{mW}$ linearly
polarized He-Ne laser at $632.8\mathrm{nm}$ (Thorlabs HRP170). The
inset shows the detailed information about the sample. Here, the
dark shadow represents the optical rail.}
\end{figure}

Our experimental setup shown in Fig.~\ref{Fig3} is similar to that
in Refs~\cite{Hosten2008,Qin2009}. A Gauss beam generated by a He-Ne
laser firstly impinges onto the HWP which is used to control the
light intensity to prevent the charge-coupled device (CCD) from
saturation. And then, the light beam passes through a short focal
length lens (L1) and a polarizer (P1) to produce an initially
linearly polarized focused beam. When the beam reaches the sample
interface, the SHEL takes place. The sample is a BK7 glass substrate
coated with a thin Ag film whose permittivity is
$\varepsilon=-18+0.5i$ at $632.8\mathrm{nm}$~\cite{Palik1998}. As
the reflected beam splits by a fraction of the wavelength, the two
components interfere destructively after the second polarizer (P2),
which is oblique to P1 with an angle of $90^{\circ}\pm\Delta$. In
our weak measurements experiment, we choose the angle
$\Delta=0.4^{\circ}$. Then we use L2 to collimate the beam and make
the beam shifts insensitive to the distance between L2 and the CCD.
Finally, we use a CCD to measure the amplified shift after L2. It
should be mentioned that the amplified factor $A_{w}$ is not a
constant, which verifies the similar result of our previous
work~\cite{Luo2011}.

\begin{figure}
\includegraphics[width=8.5cm]{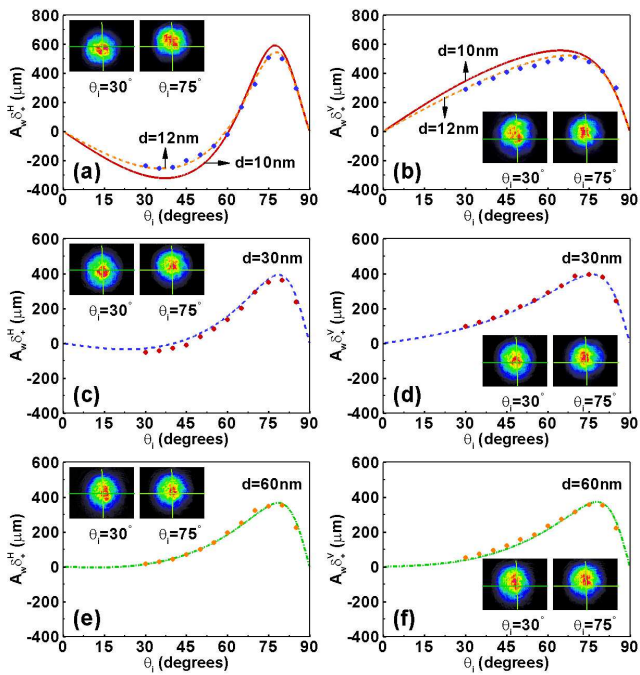}
\caption{\label{Fig4} (Color online) The amplified displacements of
horizontal polarized (left column) and vertical polarized (right
column) beam reflection on Ag film with different thicknesses:
[(a),(b)] $10\mathrm{nm}$ and $12\mathrm{nm}$, [(c),(d)]
$30\mathrm{nm}$, and [(e),(f)] $60\mathrm{nm}$. $A_{w}$ represents
the amplified factor of the weak measurements. The lines indicate
the theoretical value and the dots show the experimental results
(the error ranges are less than $10\mathrm{{\mu}m}$). The insets
show the measured field distributions from the CCD. }
\end{figure}

We measure the displacements of the SHEL on the nano-metal film
every $5^{\circ}$ from $30^{\circ}$ to $85^{\circ}$ in the case of
horizontal and vertical polarization, respectively. Limited by the
large holders of the lens, polarizers and He-Ne laser, displacements
at small incident angles were not measured. Figure~\ref{Fig4} plots
the amplified displacement in both theoretically and experimentally.
In the case of horizontal polarization, the shift first experiences
a negative value and then increases with the incident angle. After
reaching the peak value in the incident angle about $75^{\circ}$,
the shift decreases rapidly. For three different thicknesses, the
negative shifts are vary. With the thickness increasing, the range
of negative shift decreases. In the case of vertical polarization,
the shift first increases with the incident angle and also decreases
rapidly after the peak value. But, there exists no negative values
compared with the horizontal polarization.

It should be noted that the experimental results are in good
agreement with the theoretical ones when the film thicknesses are
$30\mathrm{nm}$ and $60\mathrm{nm}$ [Fig.~\ref{Fig4}(c)-(f)].
However, we observe a small deviation when the thickness is
$10\mathrm{nm}$ [Fig.~\ref{Fig4}(a) and~\ref{Fig4}(b)]. Note that
the thickness of the nano-meta film has an error in the range of
$\pm5\mathrm{nm}$, limited by the experimental condition. When the
thickness reaches to $10\mathrm{nm}$, the SHEL is very sensitive to
the error. It is the reason why there is a small deviation between
the experimental and the theoretical data. From the experimental
results, we can conclude that the actual thickness of the film is
about $12\mathrm{nm}$. This interesting characteristic may provide a
potential way for measuring the thickness of the nano-metal film.

\section{Conclusions}
In conclusion, we have observed the SHEL on a nano-metal film
experimentally via weak measurements. We have found that the
spin-orbit coupling effect can be effectively manipulated by
adjusting the thickness of the metal film. Our findings indicate
that the transverse displacement is sensitive to the thickness of
the metal film in certain range. Hence, altering the metal film
thickness will enhance or suppress the SHEL significantly. As an
analogy of spin Hall effect in an electronic system, we are able to
switch the directions of the spin accumulation in SHEL effectively
by rotating the polarization of incident light beam. These findings
provide a pathway for modulating the SHEL and thereby open the
possibility of developing nanophotonic applications.

\begin{acknowledgements}
One of the authors (X. Z.) thanks Dr. Y. Qin and Dr. N. Hermosa for
helpful discussions. We are sincerely grateful to the anonymous
referee, whose comments have led to a significant improvement on our
paper. This research was supported by the National Natural Science
Foundation of China (Grants Nos. 61025024, 11074068).
\end{acknowledgements}

\end{document}